\documentclass{article}
\usepackage[utf8]{inputenc}
\usepackage{booktabs}
\author{boj2 }
\usepackage{mathrsfs}
\usepackage{graphicx}
\usepackage{placeins}
\usepackage{float}
\graphicspath{ {./Figures/} }
\usepackage{amsmath}
\usepackage{hanging}
\usepackage{enumitem}
\usepackage{esvect}
\usepackage{setspace}
\usepackage[affil-it]{authblk}
\usepackage[left]{lineno}
\usepackage{url}
\usepackage[top=2.5cm]{geometry}
\usepackage{natbib}
\setcitestyle{authoryear,open={(},close={)}}

\vspace{0.5cm}

\begin{document}
\date{}
\title{A Land-Sea Contrast Pattern in Surface Temperature and Atmospheric Circulation Trends in Recent Decades}

\author{Benjamin O. Johnson}
\author{Maria Rugenstein}

\affil{Colorado State University}

\maketitle

\thispagestyle{plain}
\begin{abstract}
We argue land-sea contrast is an important driver of spatial patterns of observed surface temperature and atmospheric circulation trends, including the negative Inter-Decadal Pacific Oscillation (IPO)-like tendency across the Pacific basin. Observed and modeled trends display an overall decline in sea level pressure over the faster-warming land, with a spatial pattern that resembles the seasonal cycle and modeled response to land heating. Coupled historical forcing climate model simulations underestimate the land-sea warming ratio. It is only in the early response of abrupt CO\textsubscript{2} quadrupling climate model simulations that climate models are able to recreate the observed land-sea warming ratio, in which case a negative IPO-like sea level pressure trend and surface warming pattern over the Pacific comparable to observed trends are seen. We propose that some discrepancies between observed trends and those predicted by climate models are explained by the underestimation of the land-sea warming ratio by climate models.
\end{abstract}

\pagestyle{plain}
\section{Introduction}
While various anthropogenically-forced signals are emerging in observed trends \citep{shaw2024emerging,simpson2025confronting}, the physical processes shaping spatial patterns in observed trends and the causes of discrepancies with trends predicted by climate models remain poorly understood overall. Although a warming of land relative to ocean is a dominant pattern in both observed and modeled surface temperature trends, the mechanisms of which are well-studied \citep{lambert2007control,sutton2007land,byrne2013link,toda2023energy}, existing work on how this warming pattern has shaped atmospheric circulation and SST trend patterns is limited. The potential importance of this land-sea warming pattern is evidenced in idealized land heating experiments, which show land heating can substantially alter the atmospheric circulation and sea surface temperature patterns \citep{shaw2015tug,dommenget2009ocean,ackerley2016atmosphere,zhang2026remote}. Earlier work has argued modeled sea level pressure trends in the tropics are largely explained by the simulated faster warming of land \citep{bayr2013tropospheric}, but the impacts of land-sea contrast on observed trends and discrepancies with predictions from climate models are not well understood. In this work, we argue that the faster warming of land is shaping observed atmospheric circulation trends, which in turn influence SST trends. We further suggest that an underestimation of the land-sea warming ratio by climate models contributes to some discrepancies between observed climate trends and those predicted by climate models. 

 An observed, yet poorly understood, trend that has received much attention is recent years, and a major focus of this work, is the negative Inter-Decadal Pacific Oscillation (IPO)-like, often referred to as La Niña-like, trend across the Pacific basin. From 1979-2025, SSTs have warmed more slowly in the eastern Pacific compared to the west, with increased sea level pressure and anticylonic low-level atmospheric circulation trends in both the extratropical north and south Pacific straddling the tropical Pacific, a pattern that resembles the long-term fluctuations associated with the IPO \citep{power1999inter,liu2012dynamics,cane1997twentieth,england2014recent}. Although climate models predict a faint negative IPO-like trend across the Pacific basin, its magnitude is much smaller than observed \citep[e.g.,][]{wills2022systematic,coats2017simulated,seager2022persistent,olonscheck2020broad,watanabe2021enhanced,heede2023colder}. Climate model error very likely contributes to this discrepancy \citep{wills2022systematic}, bringing doubt to model projections over the Pacific basin, with implications for regional climate change and global climate sensitivity \citep{zhou2016impact,gregory2020accurately,alessi2023surface,andrews2022effect,watanabe2024possible}. Many mechanisms have been proposed to explain this observed trend and the inability of climate models to accurately capture it, such as an ocean dynamical thermostat driven by the delayed warming of upwelling water \citep{clement1996ocean,luo2017role,heede2020time,heede2021eastern}, teleconnections to the delayed warming of the Southern Ocean \citep{kang2023global,dong2022two,zheng2025relationship}, anthropogenic aerosol forcing \citep{allen2015interhemispheric,hwang2024contribution}, ozone forcing \citep{hartmann2022antarctic,dong2025stratospheric}, and too low model resolution \citep{yeager2023reduced,kang2023global}. However, there is no consensus on and no quantitative comparison of the causes. Recently, \citet{gunther2026heating} found strong land heating induces an initial equatorial east Pacific cooling in idealized model experiments. Here we further argue that the warming of land relative to ocean is an important driver of the observed negative IPO-like trend across the broader Pacific and a likely source of model error in simulating this trend.

 While Pacific basin trends are a point of emphasis in this work, we argue more broadly that land-sea contrast is an important mechanism shaping spatial patterns in observed trends globally. The structure of this paper is as follows. First, we explain how land-sea warming contrast alters the atmospheric circulation and show the seasonal cycle and modeled response to idealized land surface heating as evidence (Section \ref{subsec:LandSeaContrastDynamics}). Next, we show that this land-sea contrast pattern is seen in both modeled and observed trends since 1979 (Section \ref{subsec:Trends}). We then explain that climate models under-predict the land-sea warming ratio, and show that observed trends closely resemble the early response in abrupt CO\textsubscript{2} quadrupling (4xCO\textsubscript{2}) experiments, in which the land-sea warming ratio is similar to that observed (Section \ref{subsec:4xCO2}). We then discuss some more specific details of the atmospheric response to land heating and discuss some remaining questions on how land-sea contrast has shaped observed trends (Section \ref{sec:OpenQuestions}). Finally, we propose that an understanding of why coupled climate models underestimate the land-sea warming ratio could constrain future climate projections (Section \ref{sec:Discussion}).

\section{Data and Methods}
We compare ERA5 reanalysis \citep{hersbach2020era5} trends from 1979-2025 to CMIP6 climate model experiments. As detailed in Table S1, we use 43 CMIP6 models that have historical forcing, SSP585 emissions scenario, and abrupt 4xCO\textsubscript{2} experiments publicly available. For modeled trends, years 1979-2014 are used from coupled historical forcing experiments and extended through 2025 using SSP-585 scenario experiments (CMIP6 historical forcing experiments only span through 2014). For abrupt 4xCO\textsubscript{2} experiments, in which CO\textsubscript{2} is abruptly quadrupled from a pre-industrial control, we compare the change for the average of the first two years after carbon dioxide is quadrupled to the average over years 1900-1950 of historical forcing experiments. The first two years are chosen as a longer period (ex: the first three years after CO\textsubscript{2} quadrupling) results in a smaller land-sea warming ratio than observed trends as the land-sea warming ratio decays with time in those experiments. Land-sea warming ratios are calculated as the area-weighted mean land warming divided by ocean warming between 70° north and south. Grid points with a land fraction of less than 0.5 are counted as ocean, and land fractions greater than 0.5 are counted as land. All data is re-gridded to a common 1° grid prior to performing analyses, including land-sea warming ratio calculations.

We also compare a pre-industrial control climate model simulation to one with land albedo increased by roughly 0.1 to better understand how land heating alters the atmospheric circulation. Although the albedo change is not entirely even among land areas, the net effect is a heating of land relative to ocean. These are run with SPEAR \citep{delworth2020spear}, a Geophysical Fluid Dynamics Laboratory (GFDL) climate model at 50km atmospheric resolution and prescribed climatological, seasonally varying SSTs \citep{johnson2025exploring}. Because SSTs are kept fixed in both simulations, comparing these simulations isolates the impact of land heating relative to ocean.

\section{Results} \label{sec:Results}
\subsection{Land-Sea Thermal Contrast Alters Atmospheric Circulation} \label{subsec:LandSeaContrastDynamics}
Although the exact details of the response to land heating are more complex, as will be discussed in Section \ref{sec:OpenQuestions}, a known first-order impact of heating land relative to ocean is to induce lower sea level pressure over land and higher sea level pressure over ocean \citep[ex: ][]{rodwell2001subtropical,liu2004relationship,miyasaka2005structure,bayr2013tropospheric}. A reason to expect this effect is the lower density of warmer air via a monsoon-like mechanism: The higher temperature over land results in thermal expansion of the air column relative to ocean, leading to greater thickness between pressure levels relative to the cooler surrounding ocean, which tends to induce horizontal divergence aloft. As divergence aloft moves air from the column over land to the ocean, there is then less air mass in the column over land and more over ocean, which results in lower surface pressure of land and increased surface pressure over ocean via the hydrostatic approximation. Due to Earth's rotation and the resulting Coriolis effect, an anomalous counter-clockwise low-level flow tends to arise around the warmer land and clockwise flow over the cooler ocean in the Northern Hemisphere, with the reverse in the southern hemisphere. The anomalous surface anticyclones that tend to arise over the oceans result in relatively low surface temperatures in the eastern ocean basins along their east flanks, where induced equatorward flow has a cooling effect via both surface heat flux changes and a favorable wind stress pattern for upwelling of cooler waters from below and a cool equatorward current, with the reverse in the western ocean basin \citep{seager2003air}. 

\begin{figure}[h!]
\centering
\includegraphics[scale=0.13]{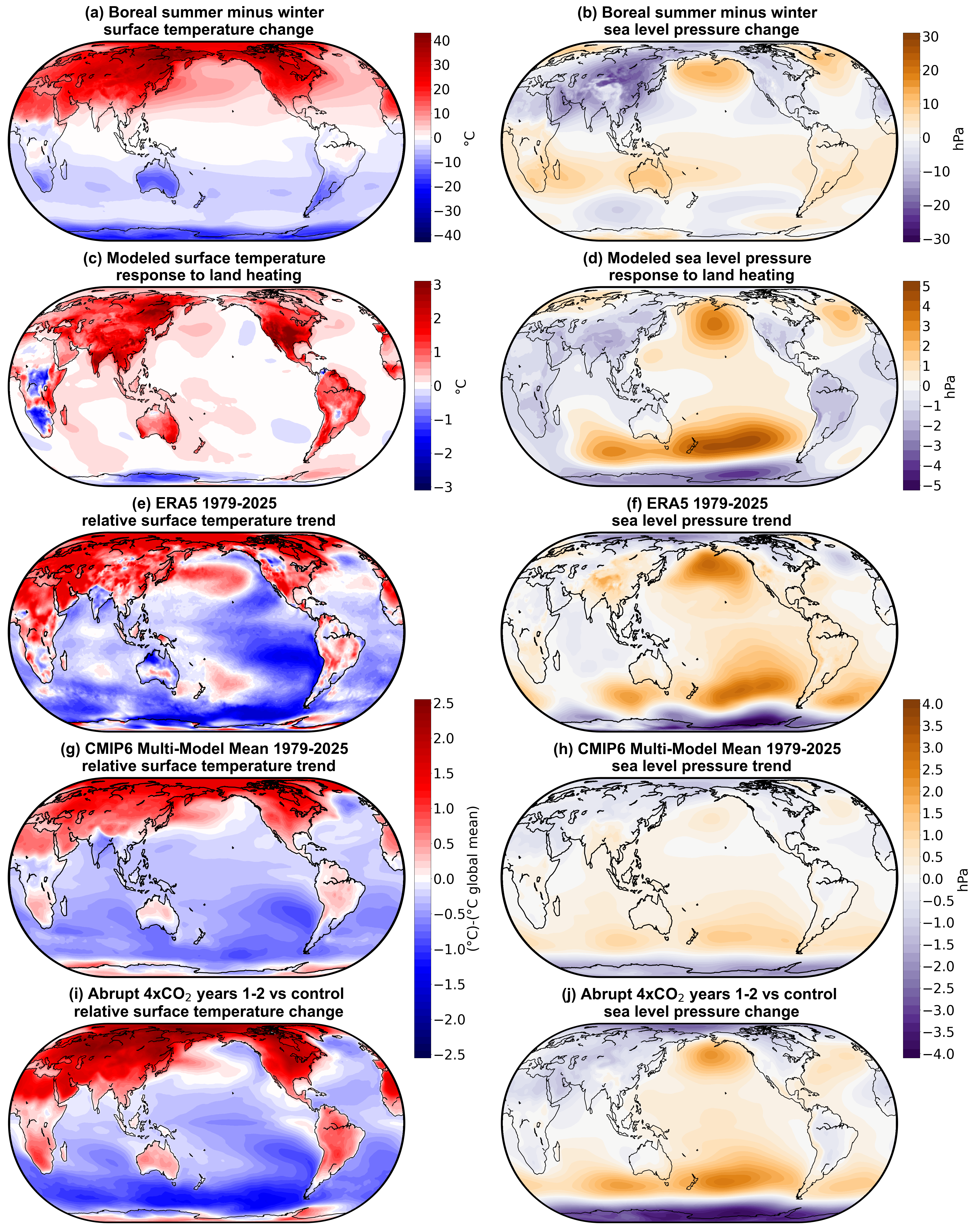}
\caption{Near-surface air temperature (left) and sea level pressure (right) change in: (a,b): The ERA5 Reanalysis seasonal cycle, defined as June-August minus December-February; (c,d): The modeled response to land heating via altered land albedo with fixed SSTs; (e,f): ERA5 Reanalysis trends from 1979-2025, (g,h): CMIP6 coupled historical forcing multi-model mean trends from 1979-2025; (i,j): The fast response of 4xCO\textsubscript{2} experiments, defined as the first two years after CO\textsubscript{2} minus the average of years 1900-1950 of historical forcing experiments. The global mean is removed in (e), (g), and (i) to emphasize the spatial pattern of change.}
\label{fig:MainFigure}
\end{figure}

This first-order impact of land-sea thermal contrast is evident in the seasonal cycle (Figure \ref{fig:MainFigure}a,b). As the northern hemisphere transitions from its cold season to warm season, land warms more than the surrounding ocean and sea level pressure declines over and near land relative to ocean, with wedges of suppressed warming along the east flanks of induced ocean surface anticyclones. Although the three-dimensional structure is more complex, the overall tendency is for anomalous ascent to arise over the land as it warms relative to ocean (Figure S1b), indicative of low-level convergence and divergence aloft over the warmer land with the reverse over the ocean, evidencing the role of the aforementioned monsoon-like mechanism. A similar but reversed pattern is seen in the southern hemisphere as it transitions to its cold season in this figure. While these changes occur alongside large changes in latent heating associated in part with seasonal monsoons, idealized model experiments, and the fact that seasonal ocean anticyclones develop gradually and not with the onset monsoon precipitation, evidence that the direct effect of surface land-sea thermal contrast dominates the seasonal pattern of sea level pressure change \citep{liu2004relationship,miyasaka2005structure,nigam2009summertime}. Although the seasonal change in sea level pressure is not exactly aligned with the temperature change pattern, evidencing the influence of complicating factors (to be discussed in Section \ref{sec:OpenQuestions}), the seasonal cycle demonstrates that a dominant effect of heating land relative to ocean is to induce lower sea level pressure over land and increased sea level pressure over ocean, an effect we will argue is evident in observed trends.

This effect of heating land relative to ocean is further evidenced in the similarity of the seasonal sea level pressure change away from the equator (where the seasonal cycle is small) to that in the modeled response to land heating via altered land albedo (Figure \ref{fig:MainFigure}c,d). This response consists of an overall sea level pressure decline and anomalous cyclonic low-level atmospheric circulation over and adjacent to the heated land and increased sea level pressure with induced anomalous anticyclonic surface flow over the surrounding ocean. Much like in the seasonal cycle, the overall tendency is for anomalous ascent to arise over the heated land (Figure S1e), evidencing the role of the monsoon-like mechanism. As we show in the following section, the spatial pattern of sea level pressure trends from 1979-2025 shares similarities with both the seasonal cycle and modeled response to land heating, suggesting that the observed stronger warming of land has shaped atmospheric circulation trends. A particularly interesting feature in the response to land heating that resembles observed trends is the negative IPO-like response in the Pacific, consisting of increased sea level pressure straddling the tropical Pacific to the north and south, and a decline in sea level pressure over the Indian Ocean relative to the Pacific, indicative of a strengthened Walker circulation. Even though SSTs are fixed in the land heating experiments, there is a slight cooling of near-surface air temperature along the east flanks of induced extratropical Pacific surface highs and increase along the west flanks, and similar fully coupled land heating experiments show an SST response that resembles the observed negative-IPO-like trend \citep{dommenget2009ocean,gunther2026heating}.

\subsection{Land-Sea Contrast in Observed and Modeled Trends} \label{subsec:Trends}
Consider ERA5 Reanalysis 1979-2025 surface temperature trends (Figure \ref{fig:MainFigure}e): A prominent pattern is that land has warmed more than ocean. In ERA5, the land-sea warming ratio between 70° north and south is 2.37, comprised of a land warming of 1.58°C and an ocean warming of 0.67°C. Along with the greater warming of land, sea level pressure has tended to decline over and downwind of land (Figure \ref{fig:MainFigure}f) in a pattern that resembles the seasonal cycle (Figure \ref{fig:MainFigure}b) and modeled response to land heating (Figure \ref{fig:MainFigure}d). For instance, the observed increase in sea level pressure over the north Pacific and increase over the subtropical north Atlantic relative to the adjacent North America and African continents are seen in both the transition from winter to summer (Figure \ref{fig:MainFigure}b) and modeled response to land heating (Figure \ref{fig:MainFigure}d). In the southern hemisphere, regions of sea level pressure increase southwest of the continents since 1979 generally map onto regions where sea level pressure increases in the transition from cold season to warm season (the reverse of Figure \ref{fig:MainFigure}a,b), and where sea level pressure increases in the modeled response to land heating. In the tropics, the decline in sea level pressure over the Indian Ocean is similar to the pattern of change in the modeled response to land heating with fixed SSTs. While there are some notable deviations from the argued land-sea contrast pattern, such as a decline in sea level pressure southeast of Greenland, and there are undoubtedly influences from additional mechanisms, it is important to note that some discrepancies may be due to internal variability imposed within the forced response, and ERA5 trends are not necessarily the ground truth. For instance, we will show in Section \ref{sec:OpenQuestions} that ground observation-based trends more closely resemble the argued land-sea contrast pattern over the subpolar north Atlantic than ERA5 trends over the same time period. 

Now consider trends predicted by CMIP6 climate models (Figure \ref{fig:MainFigure}g,h). In the shown multi-model mean, internal variability is averaged out, leaving the forced response predicted by CMIP6 climate models. In this modeled forced response, the land-sea contrast pattern is prominent. Models predict a robust warming of land relative to ocean, with a mean land-sea warming ratio of 1.78 between 70° north and south, although this ratio is significantly lower than the observed ratio (to be discussed). Consistent with the discussion in \citet{bayr2013tropospheric}, a dominant pattern in modeled trends is the tendency for sea level pressure to decline over and downstream of the faster warming land and increase over the ocean, a trend that resembles the seasonal cycle (Figure \ref{fig:MainFigure}b), modeled response to land heating (Figure \ref{fig:MainFigure}d), and observed trends (Figure \ref{fig:MainFigure}f). Interestingly, modeled trends across the Pacific basin, while of much weaker magnitude than that seen in observed trends, qualitatively resemble the observed negative IPO-like trend, with regions of slower warming in the northeast and southeast Pacific relative to further west and increased sea level pressure straddling the low-latitude Pacific. The regions of suppressed warming in the east Pacific are located near the east flanks of the regions of increased sea level pressure over the extratropical Pacific where a equatorward flow tendencies are found, similar to the pattern seen in observed trends. That land-sea warming contrast drives this faint negative IPO-like simulated trend is evidenced in  a regression of spatial trends onto the land-sea warming ratio of individual CMIP6 models, which reveals that models with a higher land-sea warming ratio tend to have a more negative IPO-like trend in the Pacific (Figure S2). The prominence of the land-sea contrast pattern in the multi-model mean historical trend, i.e. the forced response predicted by climate models, suggests that land-sea contrast is an important factor shaping the anthropogenically-forced component of observed surface temperature trends, including spatial patterns of SST trends, and atmospheric circulation trends. 

That models predict land-sea warming to have a prominent role in shaping atmospheric circulation and SST trends since 1979 is particularly interesting as they strongly underestimate the land-sea warming ratio. The modeled land-sea warming ratio is 1.78, whereas it is 2.37 in ERA5 trends. For the models used in this study, the land-sea warming ratio range from 1.34 to 2.23, land warming ranges from 0.87°C to 2.59°C, and ocean warming from 0.60°C to 1.53°C, compared to ERA5 land warming of 1.58°C and ocean warming of 0.67°C. Hence, even though some climate models capture the magnitude of observed land warming, none of them capture the observed land-sea warming ratio, suggesting that this discrepancy is at least in part a systematic bias. Given the prominence of the land-sea contrast pattern in modeled trends, it is reasonable to suspect that an underestimation of the land-sea warming ratio by climate models contributes substantially to errors in their forced response, including their much weaker negative IPO-like trend compared to observations. The following section will provide further evidence that this is the case. 

\subsection{Observed Trends Resemble the Abrupt 4xCO\textsubscript{2} Fast Response}
\label{subsec:4xCO2}
Abrupt 4xCO\textsubscript{2} experiments, in which the land-sea warming ratio is initially enhanced, provide insight into how much forcing is required for climate models to replicate the observed land-sea warming ratio. Although various feedbacks are involved, the stronger initial land-sea warming ratio is largely because after CO\textsubscript{2} is abruptly quadrupled, the ocean initially uptakes heat due to its high heat capacity whereas the land surface uptakes very little heat \citep{kamae2013tropospheric,joshi2008mechanisms,toda2023energy}. The multi-model mean land-sea warming ratio for the first two years after CO\textsubscript{2} quadrupling compared to the average over years 1900-1950 in historical forcing experiments is 2.44, similar to the ratio of 2.37 in the ERA5 1979-2025 trend. That such an extreme forcing relative to real-world anthropogenic forcing is required to reproduce the observed land-sea warming ratio evidences the significance of the discrepancy between the observed and modeled land-sea warming ratio. 

Along with a similar land-sea warming ratio as observed since 1979, the spatial pattern in the 4xCO\textsubscript{2} fast response (Figure \ref{fig:MainFigure}i,j) closely resembles the observed trend (Figure \ref{fig:MainFigure}e,f). In particular, the negative IPO-like response in the Pacific \citep[see][]{andrews2015dependence,hwang2024contribution,lu2021mechanisms} is similar to the observed trend, with increased sea level pressure in the extratropical Pacific of similar magnitude and spatial pattern as the observed trend, and regions of suppressed warming along the east flanks of the surface anticyclones where anomalous equatorward flow is induced. More broadly, sea level pressure in the lower latitudes generally increases over the Pacific basin and decreases over the far west Pacific, Indian Ocean, and Africa sector in both the early abrupt 4xCO\textsubscript{2} response and the observed trend. The close resemblance of the 4xCO\textsubscript{2} early response to observed trends over the Pacific, a pattern that is somewhat present but far too faint in modeled trends, suggests that a too-weak modeled land-sea warming ratio contributes to the inability of models to simulate the magnitude of the observed negative IPO-like trend. Indeed, as mentioned, models with a higher land-sea warming ratio tend to simulate a more negative IPO-like trend since 1979 (Figure S2). 

While SST change patterns in the early abrupt 4xCO\textsubscript{2} response are shaped in part by ocean circulation, such as a dynamic thermostat mechanism via the delayed warming of upwelled water \citep{heede2020time}, the results of \citet{gunther2026heating} suggest that the initial negative IPO-like response in abrupt 4xCO\textsubscript{2} experiments is dominated by the land-sea contrast mechanism. In their model experiments, applying 4xCO\textsubscript{2} over land only, analogous to heating land relative to ocean as in the climate model response shown in Figure \ref{fig:MainFigure}c,d, results in a negative IPO-like response similar to that seen in the general abrupt 4xCO\textsubscript{2} experiment (Figure \ref{fig:MainFigure}i). However, applying 4xCO\textsubscript{2} over only ocean yields nearly the opposite response over the Pacific (i.e., a positive IPO-like early response). As discussed in \citet{gunther2026heating}, if the ocean dynamical thermostat mechanism were dominant, one would expect an initial negative IPO-like response in the experiment with abrupt 4xCO\textsubscript{2} applied over ocean only as the delayed warming of upwelled waters would still occur. In support of this argument, we find that the change in net surface heat flux into the ocean in the early years after CO\textsubscript{2} suggests the ocean thermostat mechanism is not the dominant driver of the SST change pattern: There is generally less of an increase in ocean surface heat uptake along the east flanks of induced surface highs compared to further west (Figure S3). This suggests suppressed warming there (Figure \ref{fig:MainFigure}i) is largely due to the surface cooling effect of induced equatorward flow, a circulation response that we argue arises via the land-sea contrast mechanism. Notably, this pattern occurs despite an induced along-shore, equatorward flow component that is favorable for upwelling in the eastern ocean basins, which presumably has the effect of increasing ocean heat uptake in those areas. Thus, the faster warming of land and the resulting atmospheric circulation change shape SST patterns, which is further evidenced by the response to land heating with fixed SSTs, i.e. no ocean feedbacks, shown in Figure \ref{fig:MainFigure}c,d, which could be the case in observed trends.

\section{Open Questions} \label{sec:OpenQuestions}
While a robust direct impact of warming land relative to ocean is to induce low sea level pressure over land relative to ocean, the specifics of the response to relative land heating remain to be better understood. For instance, while the overall tendency is for anomalous ascent to develop over the heated land (Figure S1b,e,h,i), indicative of the anomalous surface convergence and divergence aloft that characterizes the monsoon-like mechanism, the spatial details are more complex. For instance, as seen in Figure S1b,e,h and depicted in Figure S4, the strongest induced subsidence over the ocean in response to land heating tends to be located in the eastern ocean basin, as noted in existing work \citep[ex: ][]{rodwell2001subtropical,liu2004relationship,miyasaka2005structure}. This pattern is overall evident in observed trends (Figure S1h) despite vertical velocity trends being more noisy than those of sea level pressure. However, there is not yet a satisfactory explanation for why in existing literature, and future work should explore the underlying cause in more detail. As depicted in Figure S4e, a likely important factor is that the induced equatorward flow there on the east flank of the induced oceanic surface anticyclone advects relatively cool air into the region, which tends to subside due to its negative buoyancy relative to the surrounding air at the same latitude, with the reverse on the west flank of the induced oceanic surface high. Induced cool air advection on the east flank of the ocean surface high also has the effect of lowering geopotential heights aloft, with the reverse on the western flank via temperature-driven changes in atmospheric thickness. Indeed, relatively low 500 hPa geopotential heights are seen in the eastern flanks in the seasonal cycle, modeled response to land heating, and observed trend (Figure S1c,f,i,l). As depicted in Figure S4d, the pattern of sea level pressure, vertical velocity and geopotential height aloft is likely also influenced by advection of the climatological wind. For instance, the the anomalous high aloft that tends to develop over the north Pacific in response to land heating may be influenced by the eastward advection of warmer continental air and warmer air due to induced low-level warm air advection in the western ocean basin. Hence, that there is not necessarily subsidence and low geopotential heights aloft over the oceanic surface high does not necessarily mean the monsoon-like mechanism is not the dominant underlying forcing mechanism of the response to relative land heating. There are numerous additional factors, such as precipitation changes, remote influences via atmospheric wave propagation, and changes in transient eddy processes (i.e. storm track changes).

\begin{figure}[h!]
\centering
\includegraphics[scale=0.13]{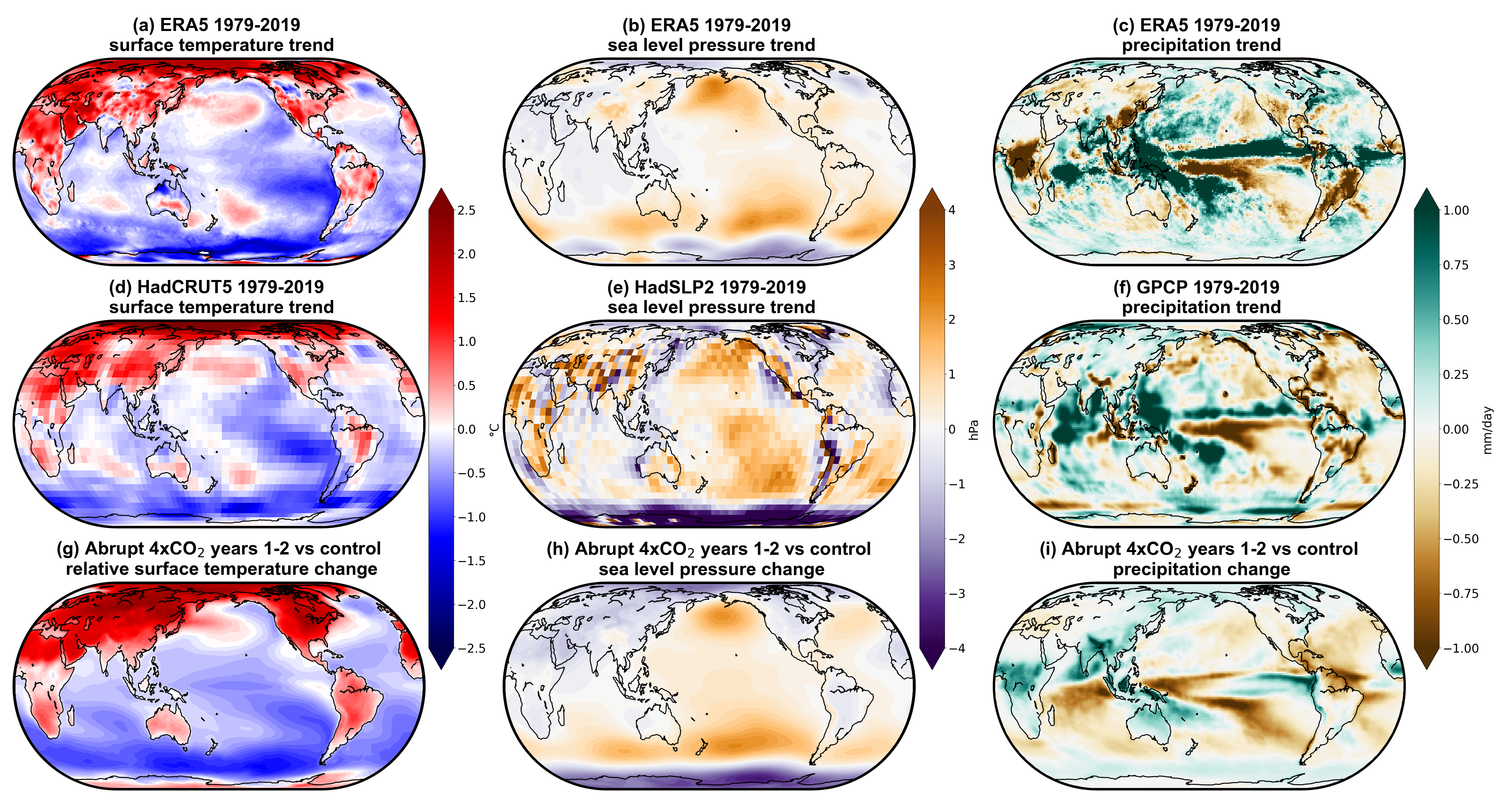}
\caption{Observational uncertainty. (a-c): ERA5 1979-2019 trends. (d): HadCRUT5 \citep{morice2021updated} 1979-2019 surface temperature trends, based on ship, buoy, and ground station observations. (e): HadSLP2 \citep{allan2006new} 1979-2019 sea level pressure trends based on ship, buoy, and ground station observations. (f): GPCP \citep{adler2003version} precipitation trends, based on satellite and ground station observations. (g-i): Abrupt 4xCO\textsubscript{2} years 1-2 minus years 1900-1950 of corresponding historical forcing experiments change.}
\label{fig:ObservationUncertainty}
\end{figure}

Additional investigation into the accuracy of observational products would also provide insight into how land-sea contrast shapes observed trends. Consider the comparison between sea level pressure trends for ERA5 (Figure \ref{fig:ObservationUncertainty}b) and Hadley Center Sea Level Pressure Dataset \citep[HadSLP2; ][]{allan2006new} (Figure \ref{fig:ObservationUncertainty}e) surface observation-based sea level pressure trends from 1979-2019 (data is only available through 2019). While the spatial pattern in trends since 1979 is similar on a global scale between ERA5 and these non-reanalysis products, there is considerable disagreement on a regional scale, and in many cases the surface observation-based sea level pressure trends are more consistent with the argued land-sea contrast pattern. For instance, an aforementioned notable deviation from the argued land-sea contrast pattern since 1979 is a decline in sea level pressure in the subpolar north Atlantic near Greenland in ERA5, which is similarly true over the shorter 1979-2019 time period. However, HadSLP2 over this period consist of a sea level pressure trend pattern over this sector that closely matches the outline of the faster-warming North American continent and areas immediately downstream (Figure \ref{fig:ObservationUncertainty}d,e), a pattern that is remarkably consistent with the argued land-sea warming contrast mechanism. While precipitation is not the focus of this work, it is noteworthy that ERA5 and GPCP \citep{adler2003version} precipitation trends also disagree substantially over this region (Figure \ref{fig:ObservationUncertainty}c,f) and many other areas. Another example is that HadSLP2 trends from 1979-2019 display a robust decline in sea level pressure west of all the subtropical continents (Figure \ref{fig:ObservationUncertainty}e), a pattern that projects onto the abrupt 4xCO\textsubscript{2} fast response (Figure \ref{fig:ObservationUncertainty}h). Such a pattern is present but less prominent in ERA5 trends over the same period (Figure \ref{fig:ObservationUncertainty}b). 

\section{Summary and Discussion} \label{sec:Discussion}
We argue land-sea warming contrast has shaped zonally asymmetric spatial patterns in atmospheric circulation trends from 1979-2025, which in turn shapes ocean warming patterns. Importantly, our analysis suggests that the faster warming of land is an important driver of the observed negative IPO-like trend across the Pacific basin, a result that is consistent with \citet{gunther2026heating}, and that an underestimation of the land-sea warming ratio in coupled historical forcing climate model experiments contributes to their inability to simulate the magnitude of this trend. We showed that a negative IPO-like response is seen in the modeled response to idealized land surface heating, and coupled historical forcing climate model experiments do qualitatively capture the observed increase in sea level pressure in the extratropical Pacific straddling the low latitudes, and suppressed surface warming along their eastern flanks where anomalous equatorward flow has a cooling effect, albeit much weaker than in observed trends. In the abrupt 4xCO\textsubscript{2} early response, where the land-sea warming ratio is similar to in the observed trend, a negative IPO-like trend comparable to observed trend than in historical forcing climate model experiments is seen \cite[see][]{andrews2015dependence,hwang2024contribution,lu2021mechanisms}.

Although historical forcing model experiments with the observed evolution of SSTs imposed do capture the IPO-like \citep{watanabe2023two,kang2025robust} sea level pressure trend in the Pacific more accurately than fully coupled experiments, our findings suggest the impacts of the observed evolution of SSTs in historical forcing experiments should be interpreted with caution. Given that the faster land warming is argued to have the direct effect of cooling the eastern ocean basin relative to the west, the observed negative IPO-like evolution of Pacific SSTs could be a response to land-sea warming contrast rather than the underlying driver itself of atmospheric circulation trends in the Pacific, with the induced SST change pattern acting as a positive feedback that further contributes to the negative IPO-like atmospheric circulation trend. Furthermore, given the global mean SST trend is too high on average in CMIP6 coupled historical forcing experiments, the ocean is forced to cool relative to land in AMIP6 experiments, resulting in a substantially increased land-sea warming ratio. For the subset of models used in this study with AMIP6 experiments available, the land-sea warming ratio from 1979-2014 between +/-70°N increases from 1.90 in CMIP6 to 2.86 in AMIP6 as the ocean warming is reduced from 0.67°C to 0.32°C. Consistent with our theory that the faster warming of land is an important driver of the negative IPO-like atmospheric circulation trend, \citet{yim2017land} finds that AMIP historical forcing experiments with a stronger land warming tend to have a relative strengthening of the Pacific Walker circulation compared to those with weaker land warming, indicative of having an amplified land-sea warming pattern. Although it is much easier to prescribe SST patterns, experiments prescribing the evolution of land surface temperature are of interest. While the historical forcing land temperature pacemaker experiments in \citet{zhang2026remote}, in which land temperature trends are nudged to be consistent with observations, do not result in a notable improvement in Pacific trends, the model used has a land warming that is too high, and thus nudging modeled land temperature trends towards observations may worsen the land-sea warming ratio issue compared to the original coupled historical forcing experiment. 

If discrepancies between modeled and observed trends are indeed linked to the land-sea warming ratio being too weak in models, we might be able to constrain future projections by determining the cause. For instance, if the discrepancy is due to an underestimation of ocean heat uptake by climate models, then the discrepancy will decay in the long term as the climate system moves towards equilibrium. Another possibility is that land moisture availability and the associated sensible to latent heat flux partition (i.e. a more positive surface evaporation trend over land in models compared to observations) is the cause of models underestimating the land-sea warming ratio. Indeed, ERA5 trends display a shift towards sensible heating over land that is much stronger than in modeled trends (Figure S5), and existing work shows the observed faster warming of land is mainly tied to a lesser increase in specific humidity over land than ocean \citep{byrne2013link,byrne2018trends} rather than a greater increase in moist static energy over land, as would be expected if the land heat flux partition were the main driver of the observed faster warming of land. Such an explanation is consistent with models failing to capture the decline in specific humidity over semi-arid regions \citep{simpson2024observed}. If the heat flux partition over land is indeed the source of the land-sea warming ratio discrepancy, there is potential for the future land-sea warming ratio to remain higher than models predict, which our findings suggest would have major implications for global and regional climate change and the prediction of it.

\section{Acknowledgements}
Benjamin O Johnson and Maria Rugenstein were supported through NOAA grant \\ NA23OAR4310596. Jackson Tobin created the schematic in Figure S4. We thank Moritz Günther, Melissa Garvais, and David Thompson for their helpful feedback. 

\section{Availability Statement}
CMIP6 data can be freely downloaded at \url{https://metagrid.esgf-west.org/search}. A list of CMIP6 models used can be found in Table S1. ERA5 near-surface air temperature, sea level pressure, and precipitation data can be freely downloaded at \url{https://doi.org/10.24381/cds.f17050d7}, and ERA5 pressure level data can be downloaded at \url{https://doi.org/10.24381/cds.6860a573}. HadSLP2 data can be freely downloaded at \url{https://www.metoffice.gov.uk/hadobs/hadslp2/data/download.html}. GPCP data can be freely downloaded at \url{https://psl.noaa.gov/data/gridded/data.gpcp.html}. Code used to make figures and calculate the land-sea warming ratios and data for the albedo experiments are available at \url{https://zenodo.org/records/20750448}. 

\section{Conflict of Interest Statement}
The authors have no conflicts of interest to disclose.

\newpage
\begin{table}[h!]
\renewcommand{\thetable}{S1}
\centering
\caption{List of CMIP6 models used in this study.}
\begin{tabular}{|l l l|}
Institution & Model & Reference\\
\hline
 CSIRO-CRCCSS & ACESS-CM2 & \cite{dix2019csiro}\\
 CSIRO-CRCCSS & ACESS-ESM1-5 & \cite{ziehn2019csiro}\\
 AWI & AWI-CM-1-1-MR & \cite{semmler2018awi}\\ 
 BCC & BCC-CSM2-MR & \cite{xin2018bcc}\\ 
 CAMS & CAMS-CSM1-0 & \cite{rong2019cams}\\ 
 CAS & CAS-ESM2-0 & \cite{chai2020cas}\\
 CCCma & CanESM5 & \cite{swart2019cccma}\\ 
 CCCma & CanESM5-1 & \cite{swart2019cccma}\\
 NCAR & CESM2 & \cite{danabasoglu2019ncar}\\
 NCAR & CESM2-FV2 & \cite{danabasoglu2019ncar}\\
 NCAR & CESM2-WACCM & \cite{danabasoglu2019ncar}\\
 CMCC & CMCC-CM2-SR5 & \cite{lovato2020cmcc}\\
 CMCC & CMCC-ESM2 & \cite{lovato2021cmcc}\\
 CNRM-CERFACS & CNRM-CM6-1-hr &\cite{voldoire2018cnrm}\\
 CNRM-CERFACS & CNRM-ESM2-1 &\cite{seferian2019cnrm}\\
 E3SM-Project & E3SM-1-0 & \cite{bader2019e3sm}\\
 EC-Earth-Consortium & EC-Earth3-CC & \cite{ec2021ec}\\
 EC-Earth-Consortium & EC-Earth3-Veg & \cite{ec2019ec}\\
 EC-Earth-Consortium & EC-Earth3-Veg-LR & \cite{ec2020ec}\\
 CAS & FGOALS-f3-L & \cite{yu2018cas}\\ 
 CAS & FGOALS-G3 & \cite{li2019cas}\\
 FIO-QLNM & FIO-ESM-2-0 & \cite{song2019ipcc}\\
 NOAA-GFDL & GFDL-ESM4 & \cite{krasting2018noaa}\\
 NASA-GISS & GISS-E2-1-G & \cite{nasa2018nasa}\\
 NASA-GISS & GISS-E2-2-G & \cite{ngifsnasa}\\
 MOHC & HadGEM3-GC31-LL & \cite{ridley2019mohc}\\
 MOHC & HADGEM3-GC31-MM & \cite{ridley2019mohc}\\
 CCCR-IITM & IITM-ESM & \cite{panickal2019cccr}\\
 INM & INM-CM4-8 & \cite{volodininm}\\
 INM & INM-CM5-0 & \cite{volodin2019inm}\\
 IPSL & IPSL-CM6A-LR & \cite{boucher2018ipcc}\\
 NIMS-KMA & KACE-1-0-G & \cite{byun2019nims}\\
 KIOST & KIOST-ESM & \cite{kim2019kiost}\\
 UA & MCM-UA-1-0 & \cite{stouffer2019u}\\
 MIROC & MIROC6 & \cite{watanabe2018miroc}\\
 MIROC & MIROC-ES2L & \cite{hajima2019miroc}\\
 MPI-M & MPI-ESM1-2-HR & \cite{jungclaus2019mpi}\\
 MPI-M & MPI-ESM1-2-LR & \cite{wieners2019mpi}\\
 MRI & MRI-ESM2-0 & \cite{yukimoto2019mri}\\
 NUIST & NESM3 & \cite{cao2019nuist}\\
 NCC & NorESM2-MM & \cite{bentsen2019ncc}\\
 AS-RCEC & TaiESM1 & \cite{lee2019rcec}\\
 MOHC & UKESM1-0-LL & \cite{tang2019mohc}\\
\bottomrule
\end{tabular}
\end{table}

\newpage
\renewcommand{\thefigure}{S\arabic{figure}}
\setcounter{figure}{0}
\begin{figure}[h!]
\centering
\includegraphics[scale=0.165]{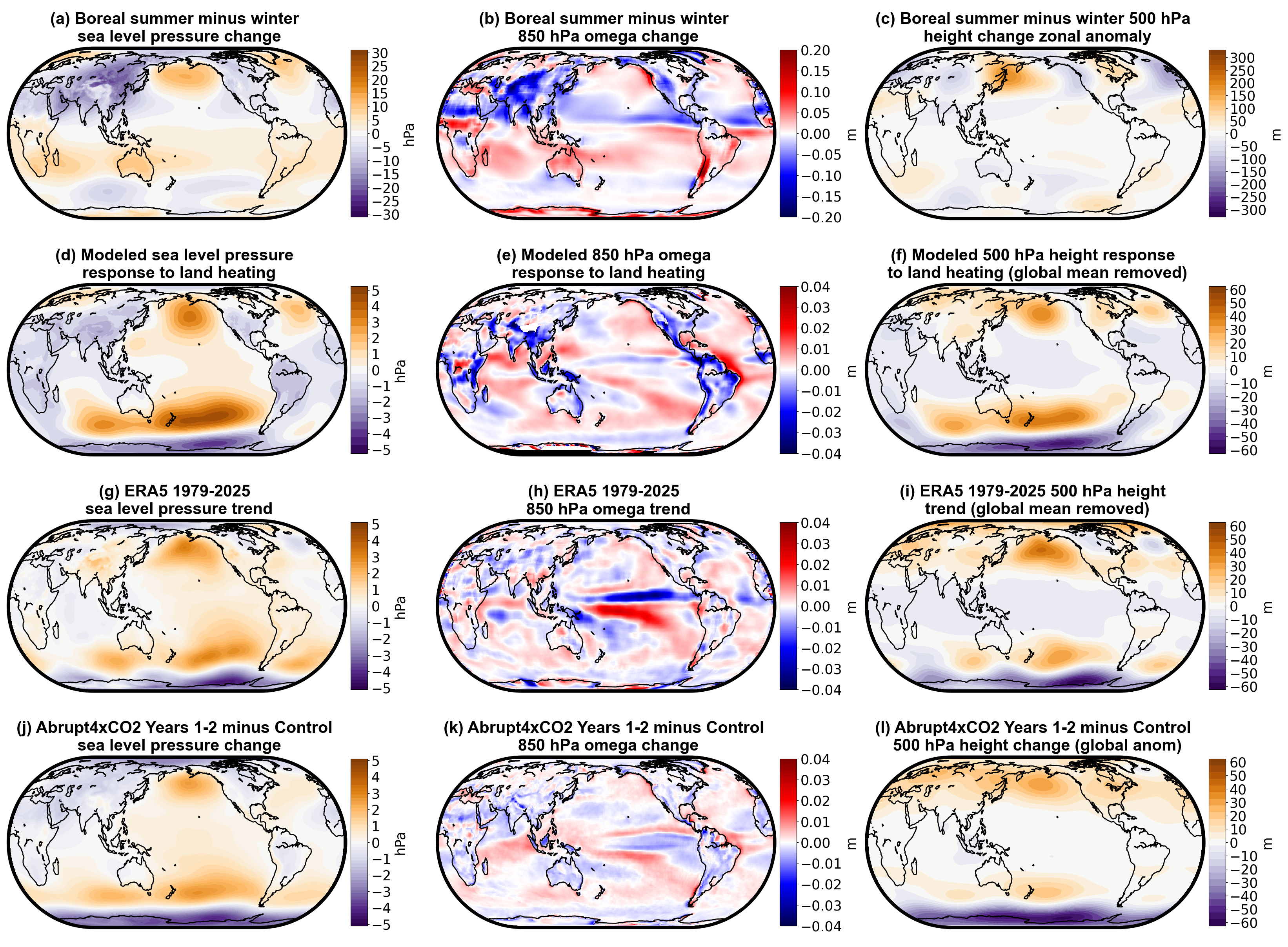}
\caption{Sea level pressure (left), 850 hPa vertical velocity (middle), and 500 hPa height (right) change in the seasonal cycle (top), modeled response to land heating (middle), ERA5 1979-2025 trend (bottom), and Abrupt 4xCO\textsubscript{2} early response. The zonal mean 500 hPa height change is removed in (c) and the global mean 500 hPa height change is removed in (f),(i),(l) to highlight spatial patterns. Here one model per institution in Table S1 is used.}
\label{fig:Z500_Omega}
\end{figure}

\newpage
\renewcommand{\thefigure}{S\arabic{figure}}
\setcounter{figure}{1}
\begin{figure}[h!]
\centering
\includegraphics[scale=0.19]{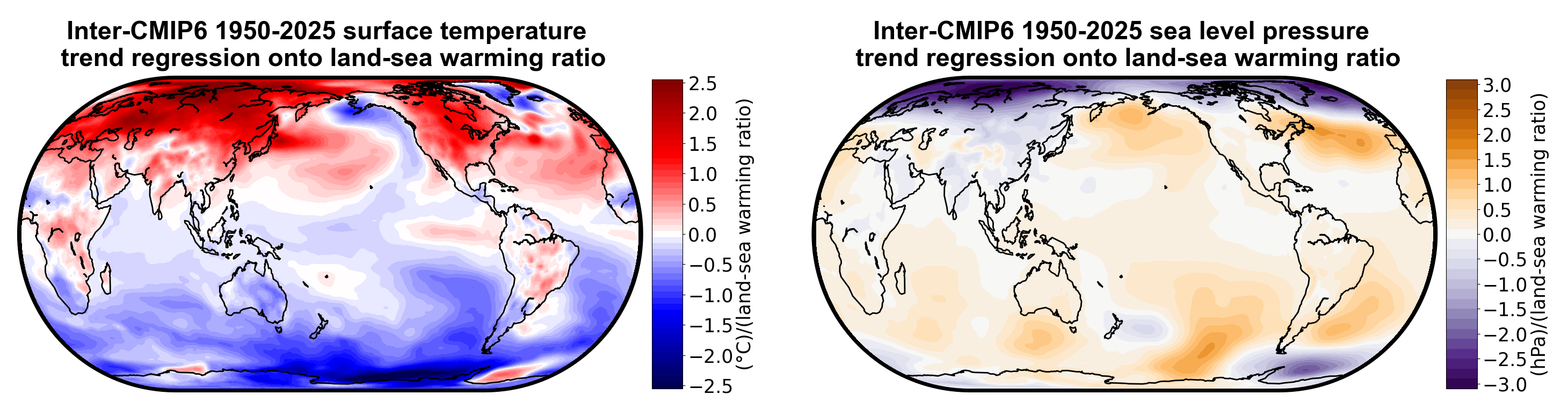}
\caption{Inter-CMIP6 1979-2025 trend regression onto land-sea warming ratio over this period. Each model's trend is divided by the global mean temperature trend before regressing with the land-sea warming ratio. The period starting at 1950 is used instead of 1979 in order to increase the signal to noise ratio.}
\label{fig:CMIP6_TrendRegression}
\end{figure}
\clearpage

\newpage
\renewcommand{\thefigure}{S\arabic{figure}}
\setcounter{figure}{2}
\begin{figure}[h!]
\centering
\includegraphics[scale=0.25]{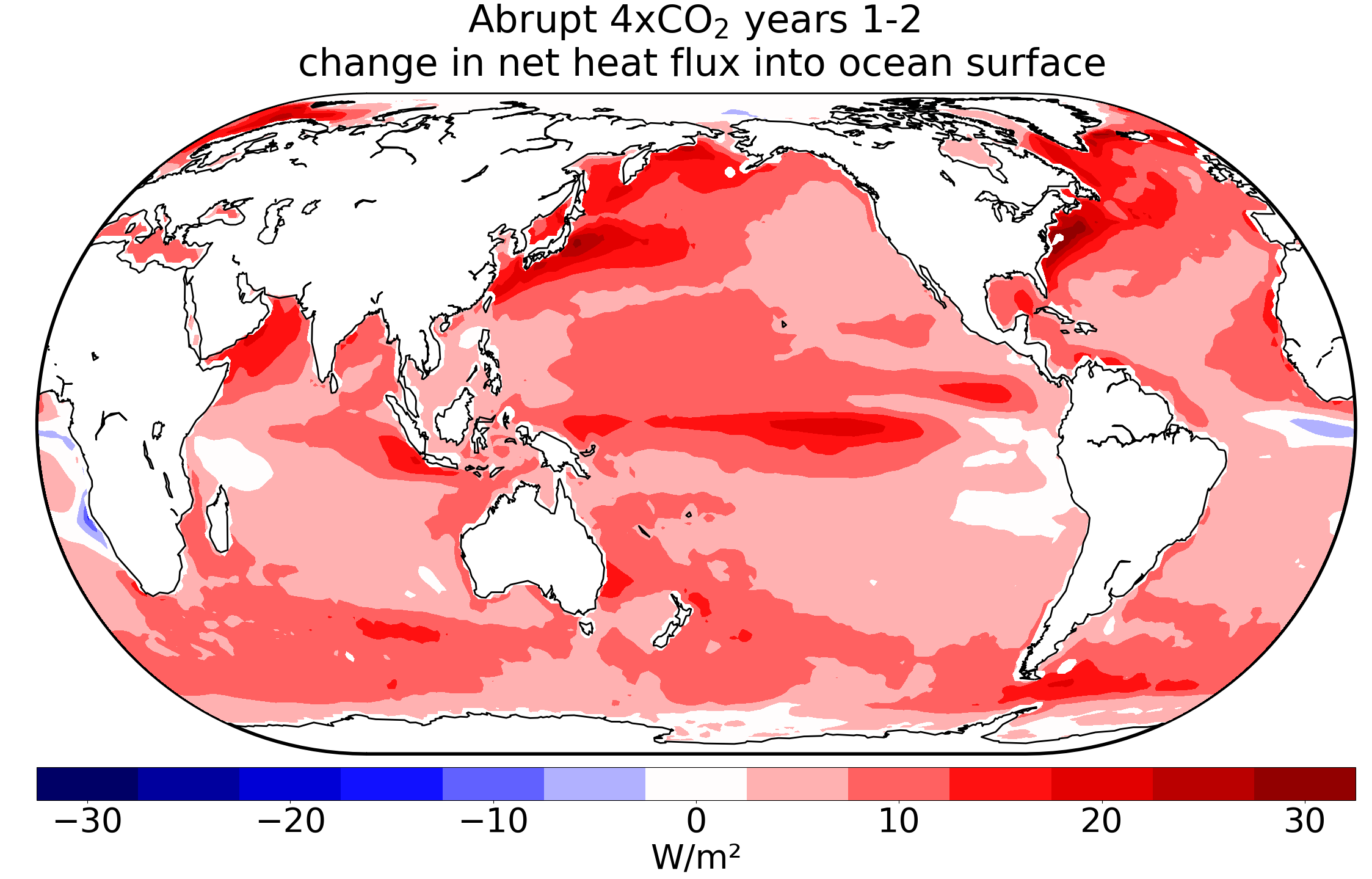}
\caption{Change in net surface heat flux into the ocean in the fast response of 4xCO\textsubscript{2} experiments, defined as the first two years after CO\textsubscript{2} minus the average of years 1900-1950 of historical forcing experiments.}
\label{fig:HeatFlux}
\end{figure}
\newpage

\renewcommand{\thefigure}{S\arabic{figure}}
\setcounter{figure}{3}
\begin{figure*}[t!]
\centering
\includegraphics[scale=0.5]{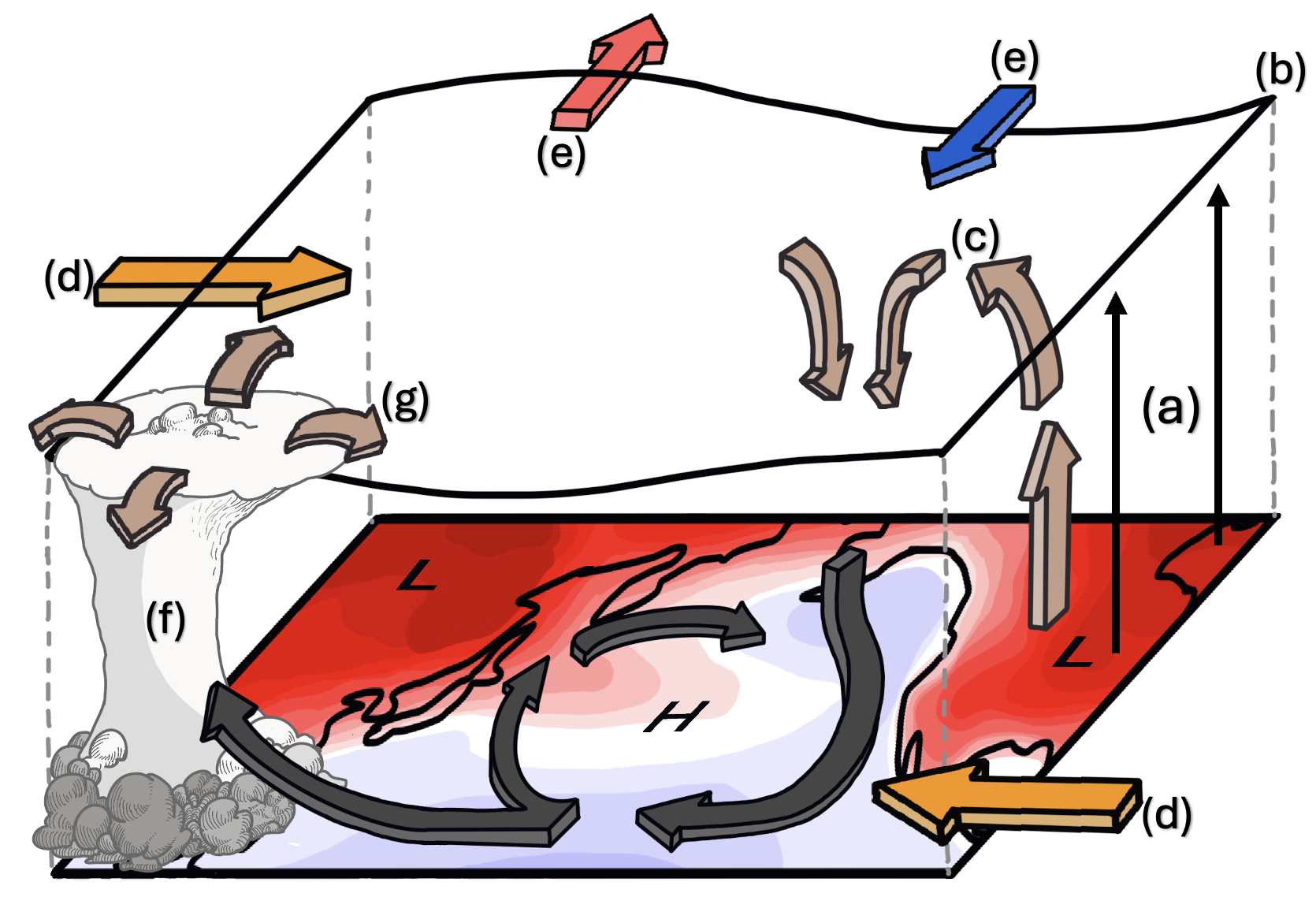}
\caption{Schematic of the response to heating land relative to ocean and some atmospheric processes that shape it. The north Pacific is illustrated, but similar processes act elsewhere. (a): thermal expansion; (c): induced horizontal divergence/convergence aloft; (d) advection of anomalies by climatological flow; (e) warm air advection by induced poleward flow and cold air advection by induced equatorward flow; (f) changes in latent heating via precipitation changes; (g) Rossby wave generation and propagation. Illustration credit: Jackson Tobin.}
\label{fig:Sketch}
\end{figure*}
\clearpage

\newpage
\renewcommand{\thefigure}{S\arabic{figure}}
\setcounter{figure}{4}
\begin{figure}[h!]
\centering
\includegraphics[scale=0.1]{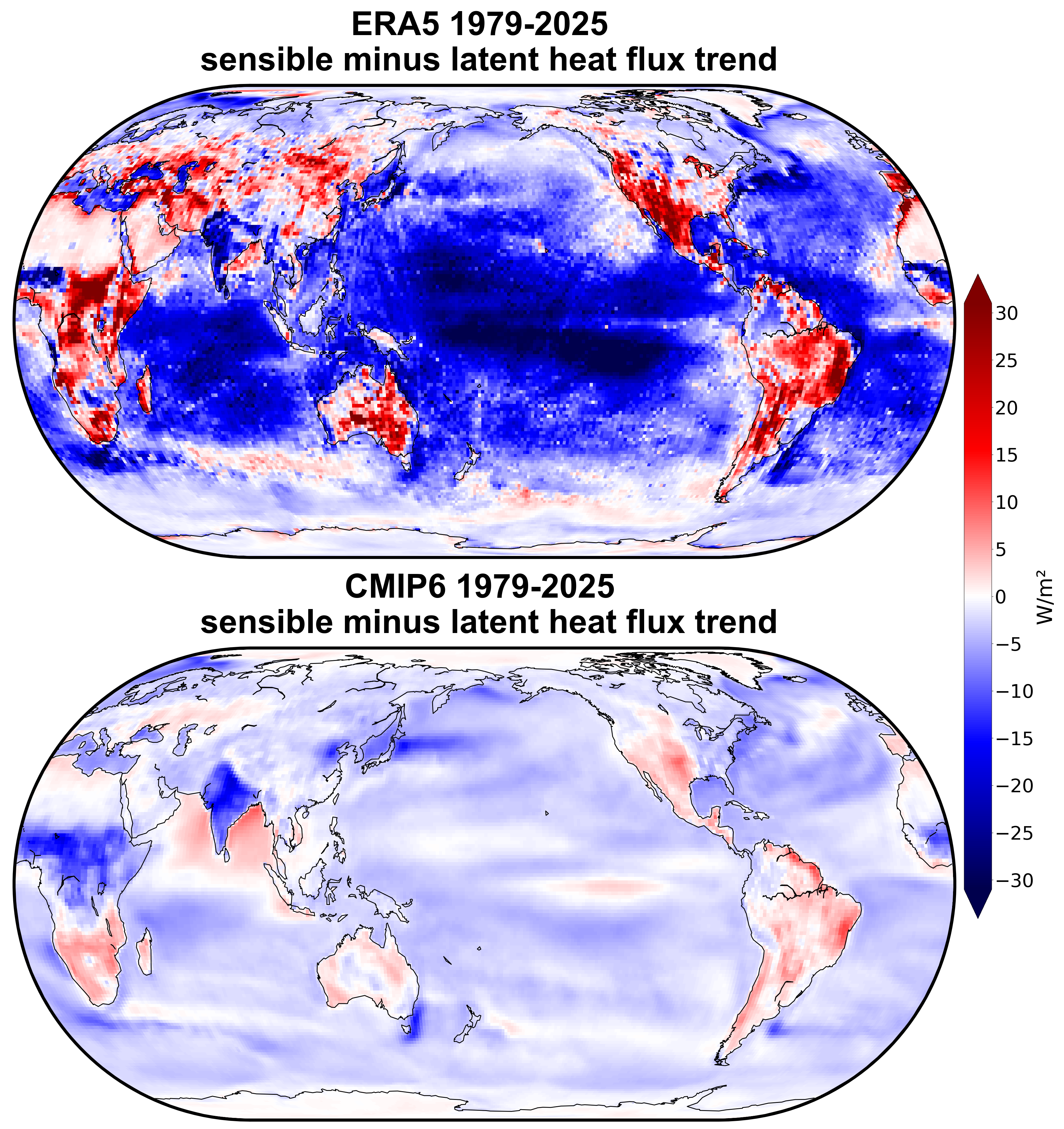}
\caption{1979-2025 trends in sensible minus latent (sensible-latent) heat flux for ERA5 (top) and CMIP6 (bottom).}
\label{fig:FluxPartition}
\end{figure}

\clearpage
\bibliography{references.bib}
\end{document}